\documentclass[aip,graphicx,reprint,jcp]{revtex4-1}

\usepackage{graphicx}
\usepackage{amsmath}
\usepackage{amssymb}
\usepackage{chemarr}
\usepackage[english]{babel}

\usepackage{color}

\draft 

\begin{document}

\title{Binding kinetics of membrane-anchored receptors and ligands: molecular dynamics simulations and theory} 

\author{Jinglei Hu}
\affiliation{Max Planck Institute of Colloids and Interfaces, Department of Theory and Bio-Systems, 14424 Potsdam, Germany}
\affiliation{Kuang Yaming Honors School, Nanjing University, 210023 Nanjing, China}
\author{Guang-Kui Xu}
\affiliation{Max Planck Institute of Colloids and Interfaces, Department of Theory and Bio-Systems, 14424 Potsdam, Germany}
\affiliation{International Center for Applied Mechanics, State Key Laboratory for Strength and Vibration of Mechanical Structures, Xi'an Jiaotong University, Xi'an 710049, China}
\author{Reinhard Lipowsky}

\affiliation{Max Planck Institute of Colloids and Interfaces, Department of Theory and Bio-Systems, 14424 Potsdam, Germany}
\author{Thomas R.\ Weikl}
\affiliation{Max Planck Institute of Colloids and Interfaces, Department of Theory and Bio-Systems, 14424 Potsdam, Germany}

\begin{abstract}
The adhesion of biological membranes is mediated by the binding of membrane-anchored receptor and ligand proteins. Central questions are how the binding kinetics of these proteins is affected by the membranes and by the membrane anchoring of the proteins.  In this article, we (i) present detailed data for the binding of membrane-anchored proteins from coarse-grained molecular dynamics simulations, and (ii) provide a theory that describes how the binding kinetics depends on the average separation and thermal roughness of the adhering membranes, and on the anchoring, lengths, and length variations of the proteins.  An important element of our theory is the tilt of bound receptor-ligand complexes and transition-state complexes relative to the membrane normals. This tilt results from an interplay of the anchoring energy and rotational entropy of the complexes and facilitates the formation of receptor-ligand bonds at membrane separations smaller than the preferred separation for binding. In our simulations, we have considered both lipid-anchored and transmembrane receptor and ligand proteins. We find that the binding equilibrium constant and binding on-rate constant of lipid-anchored proteins are considerably smaller than the binding constant and on-rate constant of rigid transmembrane proteins with identical binding domains. 
\end{abstract}

\maketitle

\section{Introduction} 

Biological processes involve the binding of molecules. These molecules are either soluble, i.e.~free to diffuse throughout intracellular compartments or extracellular spaces, or are anchored to membranes that surround cells or cellular compartments, to extracellular matrices, to cytoskeletal filaments, or to other subcellular structures \cite{Alberts08}. The anchoring to membranes is of particular importance since a large fraction of the proteins of a cell are membrane proteins \cite{Krogh01}. These membrane proteins are central for numerous biological processes but much less understood than soluble proteins since their structure and function is more difficult to assess in experiments \cite{Carpenter08}.  

Key biological processes that are mediated by the binding of membrane-anchored proteins are the adhesion of cells and the adhesion of vesicles to cells or organelles in immune responses, tissue formation, cell signaling or intracellular transport. In contrast to soluble proteins, the binding of membrane-anchored proteins poses fundamental problems that are only partially understood. 
These problems concern both the binding equilibrium  \cite{Shaw97,Zhu07,Krobath09,Wu11,Leckband12,Hu13,Wu13}  and binding kinetics \cite{Huppa10,Huang10,Robert11,Axmann12,ODonoghue13,Bihr12}  and are complicated by the fact that the binding of membrane-anchored molecules depends on the local separation of the membranes \cite{Bell84,Dembo88}.  The local separation of the membranes at the anchoring sites of apposing molecules is governed both (i) by the average separation $\bar{l}$ of the membranes, which depends on the spatial position and shape of vesicles, organelles, and cells, and (ii) by the relative membrane roughness $\xi_\perp$, which results from thermally excited shape fluctuations of the membranes \cite{Weikl04,Krobath07,Reister08,Krobath09,Reister11,Bihr12,Schmidt12,Hu13}. These shape fluctuations can be enhanced by active, ATP-driven processes arising from, e.g., the coupling of cell membranes to the cytoskeleton \cite{Gov03}. 

\begin{figure*}[tbh]
\includegraphics[width=2\columnwidth]{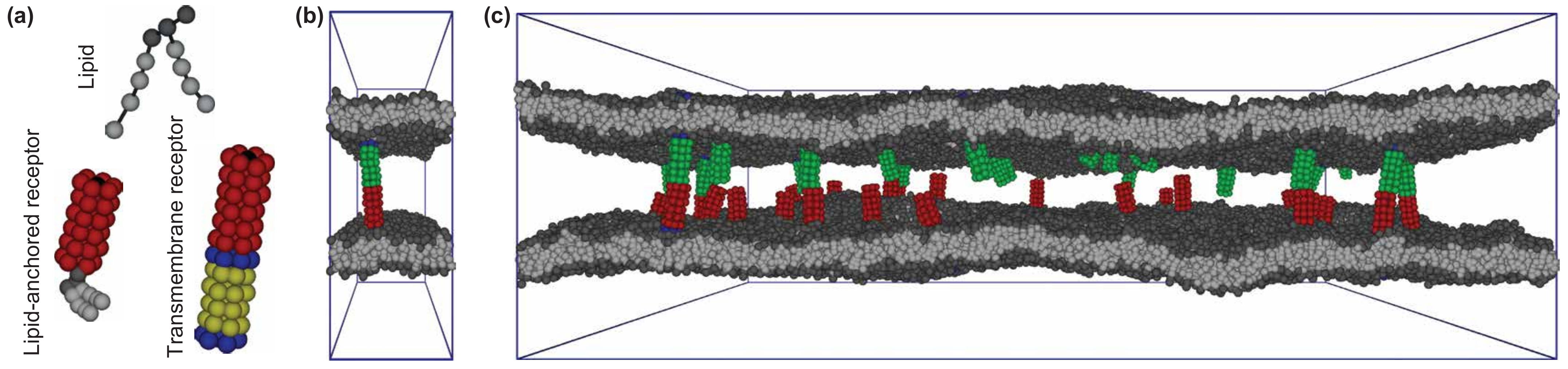}
\caption{\small (a) Coarse-grained structures of a lipid molecule, a lipid-anchored receptor, and a transmembrane receptor. The lipid molecules consist of three hydrophilic head beads (dark gray) and two hydrophobic chains with four beads each (light grey) \cite{Goetz98}. The lipid-anchored and transmembrane receptors have identical, cylindrical ectodomains that protrude out of the membranes and consists of six layers of hydrophilic beads (red), with an interaction bead or `binding site' located in the center of the top layer of beads (black) (see Appendix A for details). The ectodomain of a lipid-anchored receptor is flexibly attached to the first head bead of a lipid molecule. The ectodomain of a transmembrane receptor is rigidly connected to a cylindrical anchor composed of four  layers of hydrophobic lipid-chain-like beads (yellow) in between two  layers of lipid-head-like beads (blue). (b) Simulation snapshot of two apposing membranes with area $14\times 14$ $\text{nm}^2$ bound together by a single transmembrane receptor-ligand complex. For clarity, the ectodomain of the ligand is shown in green. Our ligands have the same structure as the receptors. (c) Simulation snapshot of two apposing membranes of area  $120\times 120$ $\text{nm}^2$ with 25 transmembrane receptors and ligands.}
\end{figure*}

In this article, we investigate detailed data for the binding kinetics and equilibrium of membrane-anchored proteins from coarse-grained molecular dynamics (MD) simulations. We have systematically varied the average separation $\bar{l}$ and relative roughness $\xi_\perp$ of the membranes in these simulations to determine the dependence of the binding rate constants $k_\text{on}$ and $k_\text{off}$ and the binding equilibrium constant $K_\text{2D}$ on $\bar{l}$ and $\xi_\perp$. Our investigation includes previous MD data for receptor and ligand proteins with transmembrane anchors \cite{Hu13} and new data for receptor and ligand proteins with lipid anchors. The transmembrane proteins and lipid-anchored proteins of our MD simulations have identical ectodomains. Differences in the binding equilibrium constants and binding on-rate and off-rate constants therefore result from differences in the anchoring. 

In addition, we present a general theory for the binding kinetics of membrane-anchored proteins that is corroborated by a comparison to our MD data. This theory is based on the general theory for the binding equilibrium constant $K_\text{2D}$ derived in our accompanying article $\cite{Xu}$ and on a general theory for the on-rate constant $k_\text{on}$ developed in this article. The off-rate constant is then obtained from the relation $k_\text{off} = k_\text{on}/ K_\text{2D}$ between the rate constants $k_\text{on}$ and $k_\text{off}$ and the binding equilibrium constant $K_\text{2D}$ of the membrane-anchored molecules. An important element of our theory is the tilt of bound receptor-ligand complexes and transition-state complexes relative to the membrane normals. This tilt results from an interplay of the anchoring energy and rotational entropy of the complexes and facilitates the formation of receptor-ligand bonds at membrane separations smaller than the preferred separation for binding.

\section{MD simulation results}

We have performed simulations of biomembrane adhesion with dissipative particle dynamics \cite{Hoogerbrugge92, Espanol95, Groot97,Shillcock02}, a coarse-grained MD technique  that explicitly includes water and reproduces the correct hydrodynamics. Coarse-grained MD simulations have been widely used to investigate the self-assembly \cite{Goetz98,Shelley01,Marrink04,Shih06} and fusion \cite{Marrink03,Shillcock05,Grafmuller07,Grafmuller09} of membranes as well as the diffusion \cite{Gambin06,Guigas06} and aggregation \cite{Reynwar07} of membrane proteins. In our simulations of biomembrane adhesion, we have considered two different types of membrane-anchored molecules: (i) receptors and ligands that are attached to a lipid molecule, and (ii) receptors and ligands with a transmembrane anchor that mimics the helix anchor of transmembrane proteins (see Fig.\ 1). These anchored receptors and ligands are free to diffuse along the fluid membranes. The specific binding of the receptors and ligands is modeled via a distance- and angle-dependent binding potential between two interaction beads located at the tip of the molecules (see Appendix A). The binding potential is identical for both types of receptors and ligands and has no barrier to ensure an efficient sampling of binding and unbinding events of receptors and ligands in our simulations. The kinetics of these events is then strongly enhanced compared with protein binding events in experiments \cite{Huppa10,Huang10,Robert11,Axmann12,ODonoghue13}. However, this rate enhancement does not affect our main results, which concern the dependence of the rate constants and equilibrium constant on the membrane separation and roughness. The binding rate constants $k_\text{on}$ and $k_\text{off}$ and binding equilibrium constant $K_\text{2D}$ can be calculated from the total numbers of binding and unbinding events observed on our simulation trajectories and from the total dwell times in the bound and unbound states of the molecules (see Appendix C). Binding and unbinding events on our simulation trajectories can be identified from the distance between the interaction beads of the molecules. 
We observe up to 10000 binding and unbinding events to determine the binding rate constants with high accuracy. In our simulations, the membranes are confined within a rectangular simulation box with size $L_x \times L_y \times L_z$ and periodic boundary conditions. Here, $L_z$ is the box extension in the direction perpendicular to the membranes, and the extensions $L_x = L_y$ specify the membrane size.{\footnote{{See supplemental movie at [URL will be inserted by AIP] for a short exemplary segment of a simulation trajectory with 15 transmembrane receptors and ligands anchored to apposing membranes of area $L_x \times L_y = 80 \times 80$ nm$^2$.}}}

\begin{figure*}[th]
\includegraphics[width=2.05\columnwidth]{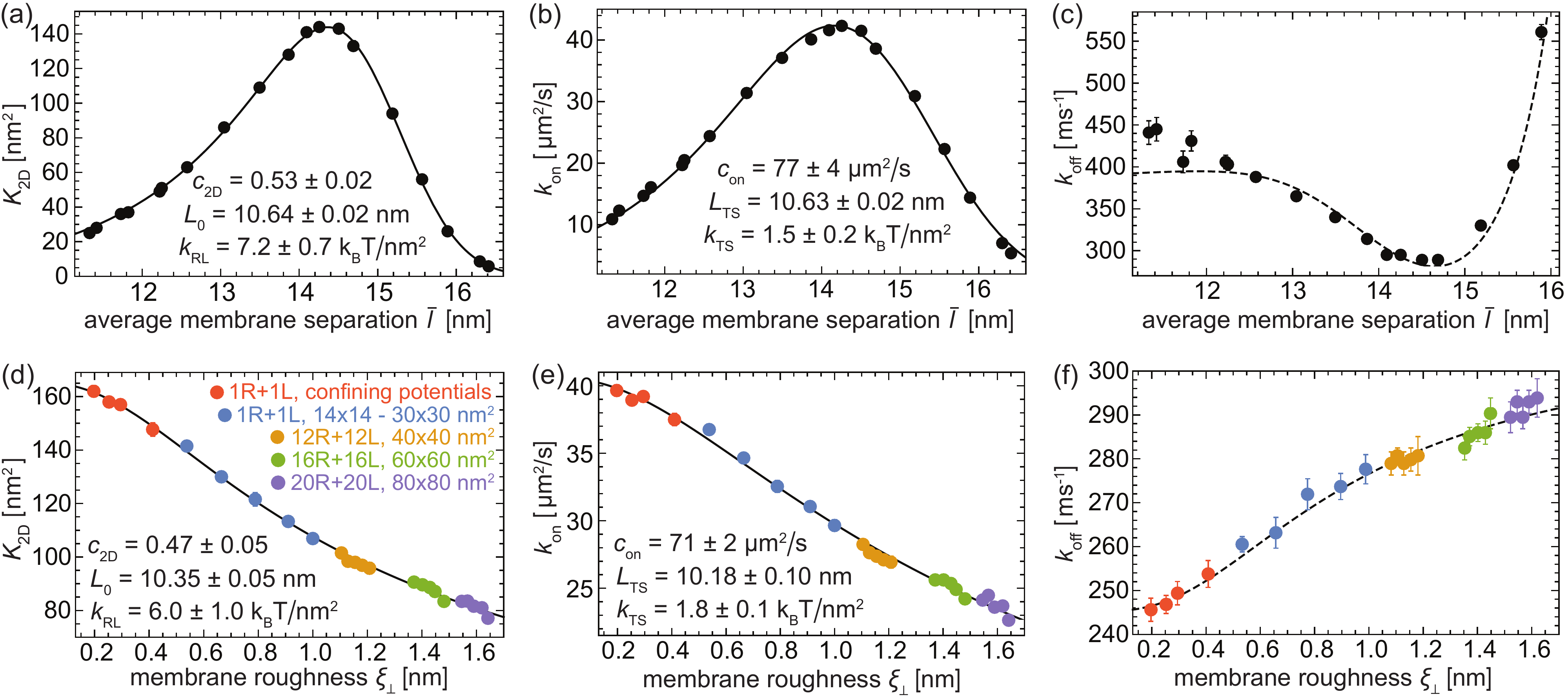}
\caption{\small Results for {\em lipid-anchored} receptors and ligands. (a) to (c): Binding equilibrium constant $K_\text{2D}$, on-rate constant $k_{\rm on}$, and off-rate constant $k_{\rm off}$ as functions of the average membrane separation $\bar{l}$ from simulations (black filled circles) with a single receptor and single ligand anchored to tensionless membranes of area $14\times 14$ nm$^2$ and relative roughness $\xi_\perp = 0.54 \pm 0.01$ nm, which is determined by the membrane area in these simulations. 
  (d) to (f): $K_\text{2D}$, $k_{\rm on}$, and $k_{\rm off}$ as functions of the relative membrane roughness $\xi_\perp$ from simulations at the preferred average membrane separation $\bar{l}_0\simeq 14.1$ nm. The red points result from simulations with single receptors and ligands and confining membrane potentials of different strength (see Appendix B for details). The other data points are obtained from simulations with tensionless membranes,  different numbers of anchored receptors and ligands, and different membrane areas. The blue data points result from simulations with membrane area $14\times 14$, $18\times 18$, $22\times 22$, $26\times 26$ and $30\times 30$ nm$^2$ (from left to right) and a single receptor and ligand.  The yellow, green, and purple data points are obtained for states with $n=1$ to 6 bound complexes of 12, 16, and 20 receptors and ligands, respectively, anchored to membranes of area $40\times 40$  nm$^2$, $60\times 60$ nm$^2$, and $80\times 80$ nm$^2$. 
  The full lines in (a), (b), (d), and (e) represent fits to Eqs.\ (\ref{K2D}) and (\ref{kon}) with fit parameters specified in the subfigures for the anchoring strength $k_a \simeq 2.5\, k_B T$ of our lipid-anchored receptors and ligands, which is obtained from the anchoring-angle distribution of unbound receptors and ligands observed in our MD simulations. The fits in (a) and (b) are obtained for the roughness $\xi_\perp \simeq 0.54$ nm, and the fits in (d) and (e) for the preferred average separation $\bar{l}\simeq 14.1$ nm.  The dashed lines in (c) and (f) follow from the full lines of (a), (b), (d), and (e) via Eq.\ (\ref{koff}). The preferred lengths $L_0$ and $L_\text{TS}$ of the lipid-anchored receptor-ligand and transition-state complexes are determined between the anchor points of the ectodomains of the receptors and ligands (see text for details). The average membrane separation $\bar{l}$ is measured between the membrane midsurfaces.
 }
\end{figure*}
\begin{figure*}
\includegraphics[width=2.05\columnwidth]{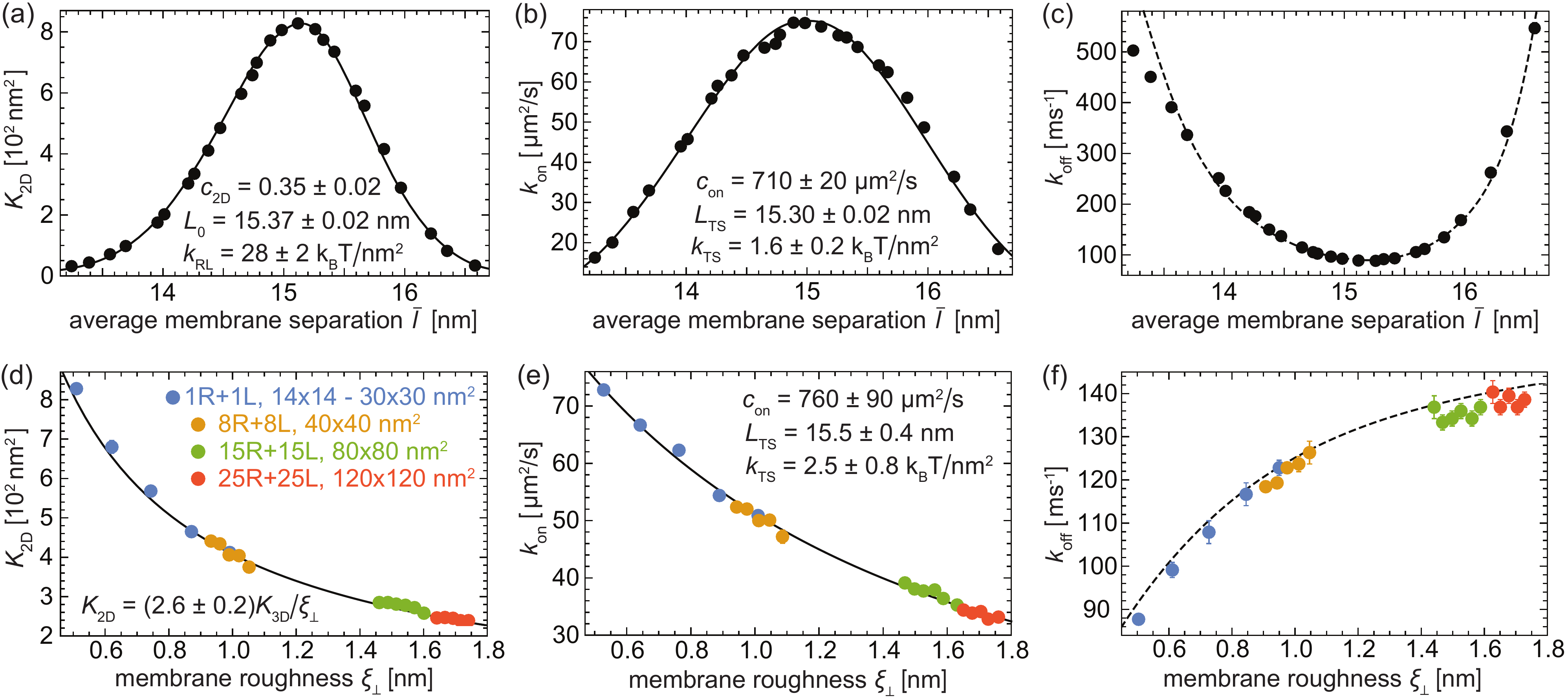}
\caption{\small Results for {\em transmembrane} receptors and ligands. (a) to (c): Binding equilibrium constant $K_\text{2D}$, on-rate constant $k_{\rm on}$, and off-rate constant $k_{\rm off}$ as functions of the average membrane separation $\bar{l}$ from simulations (black filled circles) with a single receptor and single ligand anchored to tensionless membranes of area $14\times 14$ nm$^2$ and relative roughness $\xi_\perp = 0.53 \pm 0.02$ nm, which is determined by the membrane area in these simulations. (d) to (f): $K_\text{2D}$, $k_{\rm on}$, and  $k_{\rm off}$ as functions of the relative membrane roughness $\xi_\perp$ from simulations at the preferred average separation $\bar{l}_0\simeq 15.1$ nm for binding. The data points result from simulations with tensionless membranes, different numbers of anchored receptors and ligands, and different membrane areas. The blue data points are from simulations with membrane area $14\times 14$, $18\times 18$, $22\times 22$, $26\times 26$ and $30\times 30$ nm$^2$ (from left to right) and a single receptor and ligand. The yellow data points are obtained for states with $n=1$ to 6 bound complexes of 8 receptors and 8 ligands anchored to membranes of area $40\times 40$ nm$^2$. The green data points correspond to states with $n=2$ to 8 bound complexes of 15 receptors and 15 ligands anchored to membranes  of area $80\times 80$ nm$^2$, and the red data points to states  with $n=4$ to 9 complexes of 25 receptors and 25 ligands anchored to membranes of area $120\times 120$ nm$^2$. 
The full lines in (a), (b)  and (e) represent fits to Eqs.\ (\ref{K2D}) and (\ref{kon}) with fit parameters specified in the subfigures for the anchoring strength $k_a \simeq 23\, k_B T$ of our transmembrane receptors and ligands, which is obtained from the anchoring-angle distribution of unbound receptors and ligands observed in our MD simulations. The fits in (a) and (b) are obtained for the roughness $\xi_\perp \simeq 0.53$ nm, and the fit in (e) for the preferred average separation $\bar{l}\simeq 15.1$ nm. 
The full line in (d) represents a fit to Eq.\ (\ref{K2Dlim}) for the average separation $\bar{l}=\bar{l}_0$ and $K_\text{3D}\simeq 157$ nm$^3$. The dashed lines in (c) and (f) follow from the full lines of (a), (b), (d), and (e) via Eq.\ (\ref{koff}). The preferred lengths $L_0$ and $L_\text{TS}$ of the transmembrane receptor-ligand and transition-state complexes are { determined between the centers of mass of the transmembrane anchors of the receptors and ligands located in the membrane midsurfaces}
(see text for details). 
}
\end{figure*}

The average separation $\bar{l}$ and relative roughness $\xi_\perp$ of the membranes can be varied independently in our simulations.  The average membrane separation $\bar{l}$ can be adjusted by the number of water beads between the membranes because the membranes are practically impenetrable to water beads on the simulation timescales. The relative membrane roughness $\xi_\perp$ in our simulations depends both on the size of the membranes and on the number of receptor-ligand bonds. The roughness depends on the size of the membranes because the periodic boundaries of the simulation box suppress membrane shape fluctuations with wavelength larger than $L_x/2 \pi$ where $L_x = L_y$ is the linear membrane size. In membrane systems with several anchored receptors and ligands, the roughness is affected by the number of receptor-ligand bonds because the bonds constrain the membrane shape fluctuations. To calculate the relative roughness $\xi_\perp$, we divide the x-y-plane of our simulation box, which is on average parallel to the membranes, into patches $i$ of size $2 \times 2$ nm$^2$, and determine the local separation $l_i$ of two apposing patches from the separation of the membrane midsurfaces. The roughness is then obtained as the standard deviation $\xi_\perp = \sqrt{\langle (l_i - \bar{l})^2\rangle}$ of the local separation $l_i$  from the average separation $\bar{l} = \langle l_i \rangle$ { of the membrane midsurfaces} where $\langle ... \rangle$ denotes the thermodynamic average. 

In Fig.\ 2, we present simulation results for the binding equilibrium constant $K_\text{2D}$, on-rate constant $k_\text{on}$, and off-rate constant $k_\text{off}$ of our {\em lipid-anchored} receptors and ligands. The data points in Figs.\ 2(a) to (c) result from simulations of membrane systems with different average separation $\bar{l}$ but identical roughness $\xi_\perp$. In these systems, the average separation $\bar{l}$ is varied by varying the number of water beads between the membranes, while the roughness $\xi_\perp\simeq 0.54$ nm is determined by the membrane area in these simulations. In Figs.\ 2(a) to (c), $K_\text{2D}$ and $k_\text{on}$  are maximal at a preferred average separation $\bar{l}_0$ around 14.1 nm. The data points in Figs.\ 2(d) to (f) correspond to a variety of membrane systems with an average separation $\bar{l}$ that is equal to the preferred average separation $\bar{l}_0$. The relative roughness $\xi_\perp$ of these membrane systems varies because the systems differ in membrane area, number of receptors and ligands, or membrane potential. 
{ The blue data points are from simulations of membrane systems with a single receptor and single ligand anchored to apposing membranes with various areas between $14 \times 14$ nm$^2$ and $30 \times 30$ nm$^2$ (see figure caption for details). With increasing membrane area, the relative roughness of these membrane systems increases from $\xi_\perp\simeq 0.54$ nm for the area $14 \times 14$ nm$^2$ to $\xi_\perp\simeq 1.0$ nm for the area $30 \times 30$ nm$^2$. The yellow data points in Figs.\ 2(d) to (f) are from simulations of a membrane system with 12 receptors and 12 ligands and membrane area $40 \times 40$ nm$^2$. The different values for $K_\text{2D}$ and $\xi_\perp$ in this membrane system are for states with different numbers $n$ of receptor-ligand bonds (see figure caption). These states differ in their relative membrane roughness since the receptor-ligand bonds constrain the membrane shape fluctuations. The green data points are obtained from simulations with 16 receptors and 16 ligands and membrane area $60 \times 60$ nm$^2$, and the purple data points from simulations of our largest membrane system with 20 receptors and 20 ligands anchored to apposing membranes of area $80 \times 80$ nm$^2$. Finally, the red data points are from simulations with single receptors and ligands in which the membrane fluctuations are confined by membrane potentials of different strengths. In experiments, such a situation occurs for membranes bound to apposing surfaces as, e.g., in the surface force apparatus \cite{Israelachvili92,Bayas07}. 
}
 
In analogy to Fig.\ 2, our results for membrane systems with {\em transmembrane} receptors and ligands are summarized in Fig.\ 3. The data points in Figs.\ 3(a) to (c) result from simulations of membrane systems with constant roughness $\xi_\perp\simeq 0.53$ nm, which is determined by the membrane area. The data in Figs.\ 3(d) to (f) are obtained from various systems at the preferred average separation $\bar{l}_0$ for binding, which is about 15.1 nm for our transmembrane receptors and ligands. { These membrane systems are described in detail in the caption of Fig.\ 3.}

The data in Figs.\ 2 and 3 indicate that the binding constant $K_\text{2D}$, on-rate constant $k_\text{on}$, and off-rate constant $k_\text{off}$ of membrane-anchored receptors and ligands strongly depend on the average separation $\bar{l}$ and relative roughness $\xi_\perp$ of the two apposing membranes. In addition, these data illustrate that the anchoring of the receptors and ligands strongly affects $K_\text{2D}$, $k_\text{on}$, and  $k_\text{off}$. Because our lipid-anchored and transmembrane receptor and ligand molecules have identical ectodomains, differences between the data in Figs. 2 and 3 result from the different anchoring of the molecules. Striking differences are: First,  the $K_\text{2D}$ values of our lipid-anchored receptors and ligands are substantially smaller than the $K_\text{2D}$ values of our transmembrane receptors and ligands at the preferred average membrane separation $\bar{l}_0$ for binding (see Figs.\ 2(d) and 3(d), and maxima in Figs.\ 2(a) and 3(a)). These smaller $K_\text{2D}$ values of the lipid-anchored receptors and ligands result both from smaller $k_\text{on}$ values and larger $k_\text{off}$ values, compared to the transmembrane receptors and ligands (see Figs.\ 2(e)  and (f) and 3(e) and (f)). Second, the functions $K_\text{2D}( \bar{l})$ and $k_\text{on}( \bar{l})$ for the lipid-anchored receptors and ligands shown in Figs.\ 2(a) and (b) have lower maximal values, but are broader than the corresponding functions for the transmembrane receptors and ligands shown in Figs.\ 3(a) and (b). Third, the functions $K_\text{2D}( \bar{l})$ and $k_\text{on}( \bar{l})$ are asymmetric with respect to their maxima. These asymmetries are more pronounced for the functions $K_\text{2D}( \bar{l})$ and $k_\text{on}( \bar{l})$ of our lipid-anchored receptors and ligands shown in Figs.\ 2(a) and (b).
In the next section, we introduce a general theory for the binding kinetics that helps to understand the dependence of  $K_\text{2D}$, $k_\text{on}$, and  $k_\text{off}$ both on the average separation and relative roughness of the membranes and on the anchoring of the receptors and ligands. In this theory, the asymmetries of the functions $K_\text{2D}( \bar{l})$ and $k_\text{on}( \bar{l})$ are traced back to the tilt of the receptor-ligand and transition-state complexes relative to the membrane normals.

\section{Theory}

In this section, we first consider the binding constant $K_\text{2D}(l)$ and on-rate constant $k_\text{on}(l)$ of receptors and ligands that are anchored to two planar and parallel membranes with fixed separation $l$. The local separation of such membranes is constant and identical to the membrane separation $l$. In contrast, the local separation $l$ of two apposing, fluctuating membranes varies in space and time because of thermally excited membrane shape fluctuations. We obtain  the binding constant  $K_\text{2D}$ and on-rate constant $k_\text{on}$ of receptors and ligands anchored to such fluctuating membranes  by averaging the functions $K_\text{2D}(l)$ and $k_\text{on}(l)$ over the distribution $P(l)$ of local membrane separations $l$. The mean and standard deviation of this distribution $P(l)$ are the average separation  $\bar{l}$ and relative roughness  $\xi_\perp$  of the membranes. The off-rate constant of receptors and ligands anchored to fluctuating membranes then follows from the general relation $k_\text{off} = k_\text{on}/K_\text{2D}$ between the rate constants $k_\text{on}$ and $k_\text{off}$ and binding equilibrium constant $K_\text{2D}$. 

\subsection{Binding constant of receptors and ligands anchored to planar membranes}

As shown in our accompanying article \cite{Xu}, the binding constant of rod-like receptor and ligand molecules that are anchored to two apposing planar and parallel membranes of separation $l$ can be written in the general form
\begin{equation}
K_\text{2D}(l) \simeq c_\text{2D} K_\text{3D}\frac{\sqrt{8 \pi}}{\xi_z}\frac{\Omega_\text{RL}(l)}{\Omega_\text{R}\Omega_\text{L}}
\label{K2Dl}
\end{equation}
Here, $\Omega_\text{R}$, $\Omega_\text{L}$, and $\Omega_\text{RL}(l)$ are the rotational phase space volumes of the unbound receptor R, unbound ligand L, and bound receptor-ligand complex RL relative to the membranes, $K_\text{3D}$ is the binding constant of soluble variants of the receptors and ligands without membrane anchors.  The characteristic length $\xi_z=V_b/A_b$ is the ratio of the translational phase space volume $V_b$ of the soluble RL complex and the  translational phase space area $A_b$ of the membrane-anchored complex. 
In comparison to Eq.\ (18) of our accompanying article \cite{Xu}, Eq.\ (\ref{K2Dl}) contains an additional numerical prefactor $c_\text{2D}$ to quantify deviations between theory and MD data. The rotational phase volume of an unbound receptor R or ligand L can be calculated as \cite{Xu} 
\begin{equation}
\Omega_\text{R} = \Omega_\text{L} =  2\pi \int_0^{\pi/2} e^{-\frac{1}{2} k_a \theta_a^2/k_B T} \sin \theta_a \, \text{d}\theta_a  \label{OmegaR}
\end{equation}
where $\theta_a$ is the angle between the receptor or ligand and the membrane normal, and $k_a$ is the harmonic anchoring strength. The rotational phase volume of the bound RL complex is approximately
\begin{equation}
\Omega_\text{RL}(l) \simeq 2 \pi \int_0^{\pi/2} e^{-H_\text{RL}(l,\theta_a)/k_B T}  \sin\theta_a \text{d}\theta_a
\label{OmegaRL}
\end{equation}
with the effective configurational energy 
\begin{align}
H_\text{RL}(l,\theta_a) \simeq k_a \theta_a^2 + \frac{1}{2} k_\text{RL}(l /\! \cos \theta_a - L_0)^2 
\label{HRL}
\end{align}
for binding angles and binding angle variations of the receptor and ligand in the complex that are small compared to the anchoring angle variations. The bound receptor and ligand molecules then have an approximately collinear orientation in the complex and, thus, identical anchoring angles $\theta_a$ in both membranes, with total anchoring energy $k_a \theta_a^2$. The second term of the effective configurational energy (\ref{HRL}) for the receptor-ligand complexes describes deviations from the preferred length $L_0$ of the complexes in harmonic approximation. For parallel membranes with separation $l$ and approximately identical anchoring angles $\theta_a$ of the RL complex in these membranes, the length of the complex, i.e.\ the distance between the two anchoring points in the membranes, is $L_\text{RL} \simeq l /\! \cos \theta_a$.

\subsection{On-rate constant of receptors and ligands anchored to planar membranes}

A receptor and ligand molecule can only bind at appropriate relative orientations and separations. In analogy to the Eq.\  (\ref{HRL}) for the bound receptor-ligand complex, we postulate the effective configurational energy
\begin{align}
H_\text{TS}(l,\theta_a) \simeq k_a \theta_a^2 + \frac{1}{2} k_\text{TS}(l /\! \cos \theta_a - L_{TS})^2 
\label{HTS}
\end{align}
for the transition-state complex of the binding reaction with the same anchoring strength $k_a$ as in Eq.\  (\ref{HRL}). The effective spring constant $k_\text{TS}$ for the length variations of the transition-state complex can be expected to be significantly smaller than the corresponding spring constant $k_\text{RL}$ of the RL complex, because the variations in the binding-site distance and binding angle, which affect the effective spring constants \cite{Xu}, should be significantly larger in the transition state. The preferred effective length $L_\text{TS}$ of the transition-state complex, in contrast, can be expected to be relatively close to the preferred length $L_0$ of the bound RL complex. In analogy to Eq.\ (\ref{K2Dl}) with Eq.\ (\ref{OmegaRL}), the on-rate constant is then
\begin{equation}
k_\text{on}(l) \simeq  2 \pi c_\text{on} \int_0^{\pi/2}  e^{-H_\text{TS}(l,\theta_a)/k_B T}  \sin\theta_a \text{d}\theta_a
\label{konl}
\end{equation}
for a given separation $l$ of the planar and parallel membranes. The integration over the angle $\theta_a$ in Eq.\ (\ref{konl}) can be interpreted as an integration over the transition-state ensemble of the binding reaction.

\subsection{Binding equilibrium constant and on- and off-rate constants of receptors and ligands anchored to flexible membranes}

\begin{figure}[t]
\includegraphics[width=0.95\columnwidth]{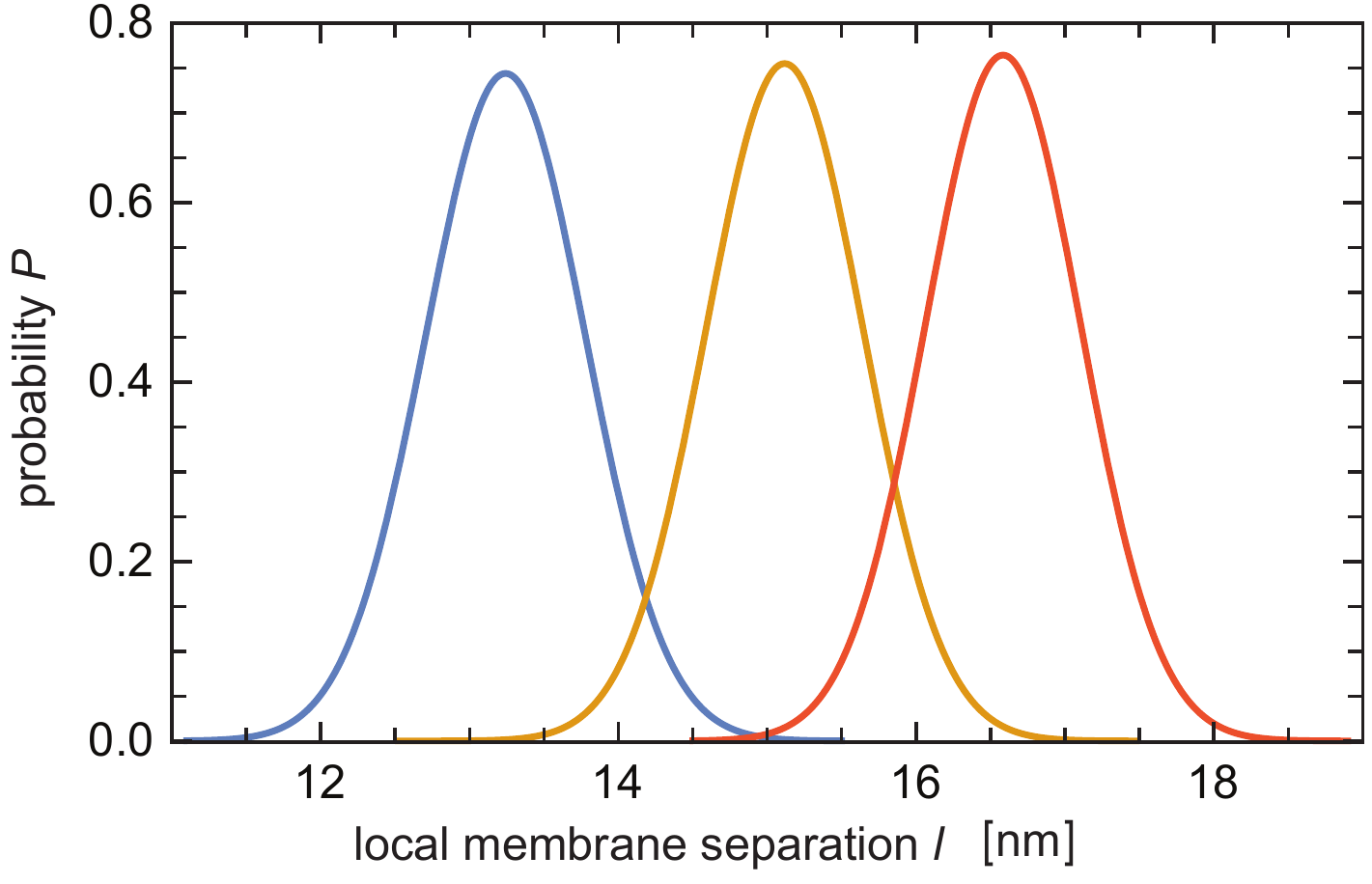}
\caption{Distributions $P(l)$ of the local membrane separation $l$ obtained from MD simulations of two apposing membranes with area $14\times 14$ nm$^2$ and single transmembrane receptor and ligand molecules at the average separations ${\bar l} = 13.24$, $15.13$ and $16.58$ nm (from left to right). The standard deviation of the distributions is the relative membrane roughness $\xi_\perp = 0.53 \pm 0.02$ nm. To calculate the local separation $l$, we divide the $x$-$y$ plane of our simulation box, which is on average parallel to the membranes, into patches of size $2\times 2$ nm$^2$, and determine the separation of the membrane midsurfaces for each pair of apposing patches. }
\label{figure_Pl}
\end{figure}

In our MD simulations, thermal fluctuations of the two apposing, flexible membranes lead to variations of the local separation $l$ of the membranes. The binding equilibrium constant can then be calculated as \cite{Xu}
\begin{equation}
K_\text{2D} = \int K_\text{2D}(l) P(l) \text{d}l
\label{K2D}
\end{equation}
where $K_\text{2D}(l)$ is given in Eq.\ (\ref{K2Dl}), and $P(l)$ is the probability distribution of the local membrane separation $l$. This probability distribution can be well approximated by the Gaussian function (see Fig.\ \ref{figure_Pl})
\begin{equation}
P(l) \simeq \exp\left[-(l-\bar l)^2/2\xi_\perp^2\right]/(\sqrt{2\pi} \xi_\perp)
\label{Pl}
\end{equation}
where $\bar{l}=\langle l \rangle$ is the average separation of the membranes or membrane segments, and $\xi_\perp = \sqrt{\langle(l - \bar{l})^2\rangle}$ is the relative roughness of the membranes. For a relative membrane roughness $\xi_\perp$ that is much larger than the width $\xi_\text{RL}$ of the function $K_\text{2D}(l)$, Eq.\ (\ref{K2D}) simplifies to \cite{Xu}
\begin{equation}
K_\text{2D} \simeq \frac{K_\text{3D} k_a }{\sqrt{2\pi k_B T k_\text{RL}}\, \xi_z\xi_\perp} e
^{-(\bar{l}_0 -\bar l)^2/2\xi_\perp^2}
\label{K2Dlim}
\end{equation}
for anchoring strengths $k_a \gg k_B T$, where $\bar{l}_0$ is the preferred average separation of the receptor-ligand complexes for large membrane roughnesses. The width of the function $K_\text{2D}(l)$ is approximately $\xi_\text{RL}\simeq \sqrt{(k_B T/k_\text{RL}) + (k_BT L_0/2k_a)^2}$ for large anchoring strengths $k_a \gg k_B T$ according to Eq.\ (23) of our accompanying article \cite{Xu}.

In analogy to Eq. (\ref{K2D}), the on-rate constant of receptors and ligands anchored to fluctuating membranes can be calculated as 
\begin{equation}
k_\text{on} = \int k_\text{on}(l) P(l) \text{d}l
\label{kon}
\end{equation}
with $k_\text{on}(l)$ given in Eq.\ (\ref{konl}). In contrast to Eq.\ (\ref{K2D}), the average over local separations in Eq.\ (\ref{kon}) for the on-rate constant $k_{\rm on}$ relies on characteristic timescales for membrane fluctuations that are significantly smaller than the timescales for the diffusion of the anchored molecules on the relevant length scales. The average in Eq.\ (\ref{K2D}) for the binding constant $K_\text{2D}$ is independent of these timescales because $K_\text{2D}$ is an equilibrium quantity that does not depend on dynamic aspects. The Eqs.\ (\ref{K2D}) and (\ref{kon}) imply spacial and temporal averages because the distribution $P(l)$ in our theory reflects variations of the local separation $l$ both in space and time. For the on-rate constant $k_\text{on}$, a related temporal average at fixed membrane locations has been previously employed by Bihr et al.\ \cite{Bihr12}.

For a relative membrane roughness $\xi_\perp$ that is much larger than the width of the function $k_\text{on}(l)$, the distribution $P(l)$ is nearly constant over the range of local separations $l$ for which $k_\text{on}(l)$ is not negligibly small. In analogy to Eq.\ (\ref{K2Dlim}), the average over local separations in Eq.\ (\ref{kon}) then simplifies to 
\begin{align}
k_\text{on}& \simeq P(\bar{l}_\text{TS}) \int k_\text{on}(l) \text{d}l  
\label{aux1}\\ 
&\simeq\frac{c_\text{on} \pi k_B T}{\sqrt{k_a k_\text{TS}}\xi_\perp}  F_D\left(\sqrt{\frac{k_B T}{k_a}}\right)e
^{-(\bar{l}_\text{TS} -\bar l)^2/2\xi_\perp^2} \\
&\simeq\frac{c_\text{on}\pi (k_B T)^{3/2} }{k_a\sqrt{k_\text{TS}}\xi_\perp} e
^{-(\bar{l}_\text{TS} -\bar l)^2/2\xi_\perp^2} \text{~~for~} k_a \gg k_B T
\label{konlim}
\end{align}
where $\bar{l}_\text{TS}$ is the preferred average membrane separation of the transition-state complex for large membrane roughnesses, and $F_D$ is the Dawson function. The integral of the function $k_\text{on}(l)$ in Eq.\ (\ref{aux1}) can be determined from Eq.\ (C3) of our accompanying article \cite{Xu}. In analogy to the width $\xi_\text{RL}$ of the function $K_\text{2D}(l)$ given below Eq.\ (\ref{K2Dlim}), the width of the function $k_\text{on}(l)$ can be estimated as $\xi_\text{TS}\simeq \sqrt{(k_B T/k_\text{TS}) + (k_BT L_\text{TS}/2k_a)^2}$ for large anchoring strengths $k_a \gg k_B T$. 

From the Eqs.\ (\ref{K2D}) and (\ref{kon}), we obtain 
\begin{equation}
k_\text{off} = \frac{k_\text{on}}{K_\text{2D}} = \frac{\int k_\text{on}(l) P(l) \text{d}l}{\int K_\text{2D}(l) P(l) \text{d}l}
\label{koff}
\end{equation}
as general result for the off-rate constant of receptors and ligands anchored to fluctuating membranes. From the Eqs.\ (\ref{K2Dlim}) and (\ref{konlim}), we obtain 
\begin{equation}
k_\text{off} \simeq \frac{\sqrt{2}\pi^{3/2} (k_B T)^2c_\text{on}\sqrt{k_\text{RL}}\, \xi_z }{k_a^2\sqrt{k_\text{TS}}K_\text{3D}} 
\label{kofflim}
\end{equation}
in the limit of large membrane roughnesses for a preferred average membrane separation $\bar{l}_\text{TS}$ of the transition-state complex that is approximately equal to the preferred average separation $\bar{l}_0$ of the bound receptor-ligand complex. The limiting value of $k_\text{off}$ in Eq.\ (\ref{kofflim}) is independent both of the average membrane separation $\bar{l}$ and the relative membrane roughness $\xi_\perp$. 

\section{Comparison of theory and MD results}

\begin{figure}[t]
\includegraphics[width=0.95\columnwidth]{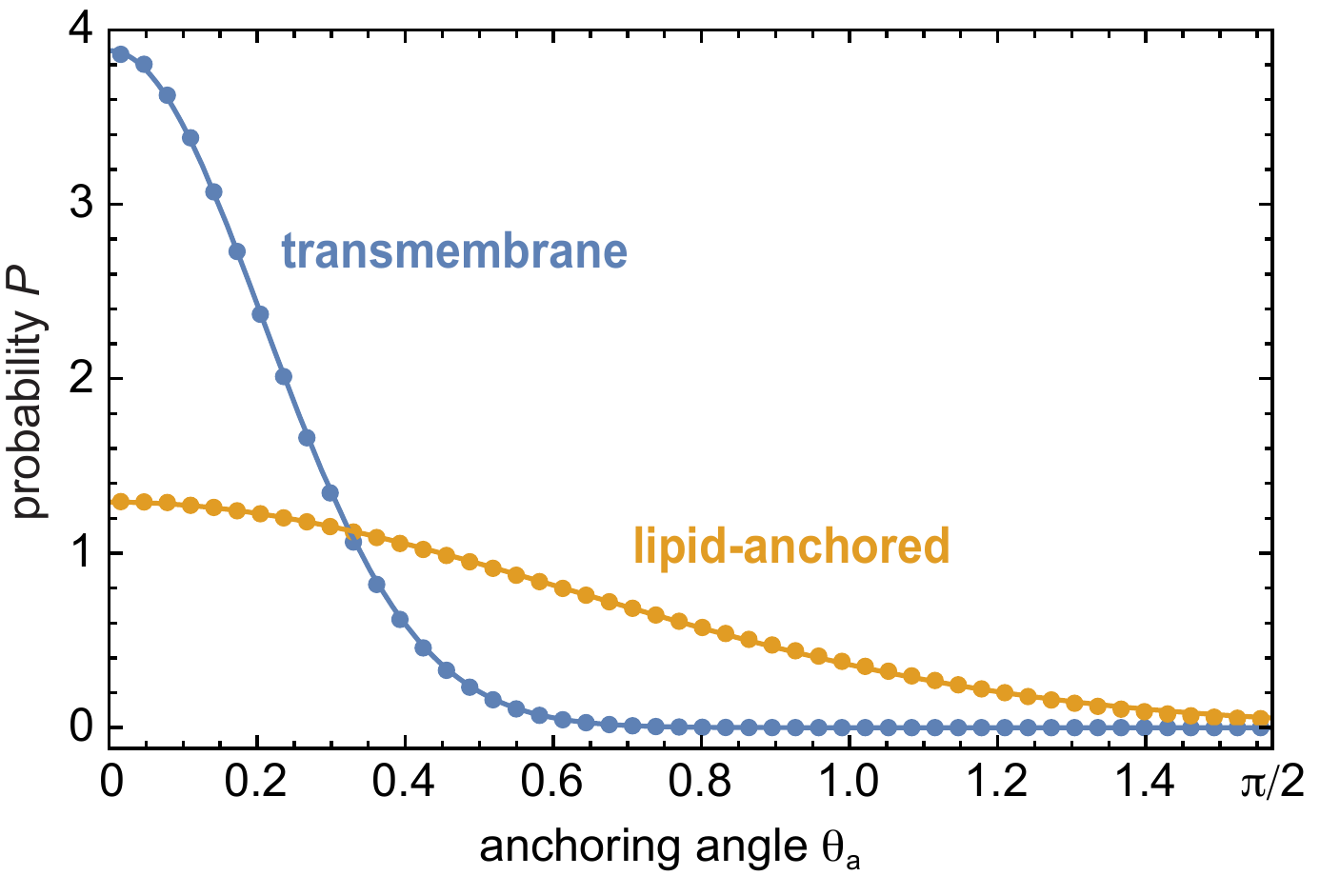}
\caption{
Rescaled probability distributions $P(\theta_a)$ of the anchoring angles $\theta_a$ of unbound transmembrane and lipid-anchored receptors and ligands obtained from MD simulations of two apposing membranes with area $14\times 14$ nm$^2$  and a confining membrane potential of strength $k_{\rm conf} = 64\, k_BT$/\text{nm}$^2$ (see Appendix B for details). The lines represent fits $P(\theta_a) \sim \exp(-\frac{1}{2}k_a \theta_a^2/k_B T)$ to the MD data points with the values $k_a\simeq 2.5\, k_B T$ for the anchoring strength of lipid-anchored receptors and ligands and $k_a\simeq 23\, k_B T$ for transmembrane receptors and ligands. The rescaled distributions $P(\theta_a)$ are derived from the anchoring-angle distributions $P(\theta_a)\sin\theta_a$ observed in the MD simulations.
}
\label{figure_anchoring_angles}
\end{figure}

We now compare our general theoretical results for the binding constant $K_\text{2D}$, on-rate constant $k_\text{on}$, and off-rate constant $k_\text{off}$ of receptors and ligands that are anchored to fluctuating membranes with our MD data. Two important aspects for this comparison are: First, the anchoring strength $k_a$ of the lipid-anchored and transmembrane receptors and ligands of our MD simulations can be estimated from the anchoring-angle distributions of the unbound receptors and ligands shown in Fig.\ \ref{figure_anchoring_angles}. From these distributions, we obtain the anchoring strength $k_a \simeq 2.5\, k_B T$ for our lipid-anchored receptors and ligands, and the anchoring strength $k_a \simeq 23 \, k_B T$ for our transmembrane receptors and ligands. Second, our lipid-anchored receptors and ligands tilt at anchoring points that are separated by a distance $l_a$ from the membrane midsurfaces between which the membrane separation is measured in our simulations. We therefore substract a value $2l_a$ from the MD values for the average membrane separation $\bar{l}$ given in Fig.\ 2 in the comparison to our theoretical results. We obtain the estimate $2l_a  \simeq d_\text{HH} + l_0 \simeq 4.5$ nm from the value $d_\text{HH}\simeq 3.64$ nm for the vertical distance between the head-group layers of a membrane (see appendix B) and the value $l_0 = 0.875$ nm for the preferred length of the bond that connects the lipid head to the ectodomain of a lipid-anchored receptor or ligand (see appendix A). We take the anchoring point of a lipid-anchored receptor or ligand to be at the center of this bond. Our transmembrane receptors and ligands are rather rigid and tilt as whole molecules relative to the membrane. We therefore take the anchoring points of our transmembrane receptors and ligands to be located at the membrane midsurfaces, { i.e.\ in the centers of mass of the transmembrane anchors of the receptors and ligands.}  

Our general theoretical result for the binding constant $K_\text{2D}$ given in Eq.\ (\ref{K2D}) with Eqs.\ (\ref{K2Dl}) to (\ref{HRL}) and (\ref{Pl}) includes the anchoring strength $k_a$ of the receptors and ligands, the preferred length $L_0$ of the receptor-ligand complex, the effective spring constant $k_\text{RL}$ for length variations  of the complex, the length scale $\xi_z$ and the binding constant $K_\text{3D}$ of soluble variants of the receptors and ligands as characteristic parameters of the receptor and ligand molecules. In addition, this general equation for $K_\text{2D}$ includes the average separation $\bar{l}$ and relative roughness $\xi_\perp$ in Eq.\ (\ref{Pl}) as characteristic parameters of the membranes, and contains a numerical prefactor $c_\text{2D}$ in Eq.\ (\ref{K2Dl}). The full lines in the Figs.\ 2(a) and (c) represent fits of Eq.\ (\ref{K2D}) to the MD data for the anchoring strength $k_a \simeq 2.5\, k_B T$ of our lipid-anchored receptors and ligands and the values $\xi_z \simeq 0.19$ nm and $K_\text{3D}\simeq 157$ nm$^3$ obtained from MD simulations \cite{Hu13}. The values of the three fit parameters $c_\text{2D}$, $L_0$, and $k_\text{RL}$ are specified in the figures. Similarly, the full line in Fig.\ 3(a) results from a  fit of Eq.\ (\ref{K2D}) for the anchoring strength  $k_a \simeq 23\, k_B T$ of our transmembrane receptors and ligands, $\xi_z \simeq 0.19$ nm, and $K_\text{3D}\simeq 157$ nm$^3$. The full line in Fig.\ 3(d) represents a fit to Eq.\ (\ref{K2Dlim}) for $K_\text{2D}$ in the limit of large roughnesses $\xi_\perp$ for the anchoring strength $k_a \simeq 23\, k_B T$  and average separations $\bar{l}$ that are equal to the preferred average separation $\bar{l}_0$ of the bound complex. We fit here to Eq.\ (\ref{K2Dlim}) because the relative roughnesses $\xi_\perp$  of Fig.\ 3(d) are significantly larger than the width $\xi_\text{RL}$ of the function $K_\text{2D}(l)$ of our transmembrane receptors and ligands. Based on Eq.\ (23) of our accompanying article \cite{Xu}, the width $\xi_\text{RL}$ of the function $K_\text{2D}(l)$ can be estimated as $\xi_\text{RL}\simeq \sqrt{(k_B T/k_\text{RL}) + (k_BT L_0/2k_a)^2} \simeq 0.38$ nm for $k_a \simeq 23\, k_B T$ and the fit values $L_0 \simeq 15.37$ nm and $k_\text{RL}\simeq 28 \, k_B T/\text{nm}^2$ of the transmembrane receptors and ligands from Fig.\ 3(a).

Our general theoretical result for the on-rate constant $k_\text{on}$ given in Eq.\ (\ref{kon}) with Eqs.\ (\ref{HTS}), (\ref{konl}) and (\ref{Pl}) includes the anchoring strength $k_a$ of the receptors and ligands, the preferred length $L_\text{TS}$ of the transition-state complex, and the effective spring constant $k_\text{TS}$ for length variations of the transition-state complex as characteristic parameters of the receptors and ligands, besides the prefactor $c_\text{on}$ in Eq.\ (\ref{konl}) and the average separation $\bar{l}$ and relative roughness $\xi_\perp$ of the apposing membranes. The full lines in the Figs.\ 2(b) and (e) and 3(b) and (e) represent fits of Eq.\ (\ref{kon}) to the MD data for the anchoring strengths $k_a \simeq 2.5\, k_B T$  and $k_a \simeq 23\, k_B T$ of our lipid-anchored and transmembrane receptors and ligands, respectively. The values of the three fit parameters $c_\text{on}$, $L_\text{TS}$, and $k_\text{TS}$ are specified in the figures. The dashed lines in the Figs.\ 2(c) and (f) and 3(c) and (f) for the off-rate constant are obtained from the full lines of the other subfigures via the general relation $k_\text{off} = k_\text{on}/ K_\text{2D}$ between the rate constants $k_\text{on}$ and $k_\text{off}$ and the binding equilibrium constant $K_\text{2D}$. 

The full lines in the Figs.\ 2 and 3 obtained from fits of our general theoretical results for $K_\text{2D}$ and $k_\text{on}$ are in good agreement with the MD data. From Fig.\ 3(a), we obtain the fit value $k_\text{RL} = 28 \pm 2 \, k_B T/\text{nm}^2$ for the effective spring constant for length variations of the bound complex of our transmembrane receptors and ligands. This fit value is in agreement with the value $k_BT/\xi_z^2 \simeq 28 \, k_B T/\text{nm}^2$ obtained from Eq.\ (15) of our accompanying article \cite{Xu} for $\xi_z \simeq 0.19$ nm and $\sigma_b L^2/4\ll \xi_z$, which holds for the standard deviation $\sigma_b = 0.084$ of the binding-angle distribution of the bound complex \cite{Hu13} and the length $L \simeq 7.4$ nm of the transmembrane receptors and ligands measured from the membrane midsurfaces. This agreement indicates that the effective spring constant $k_\text{RL}$ for length variations of our transmembrane receptor-ligand complex is dominated by the standard deviation $\xi_z$ of the distance between the two binding sites in the direction of the receptor-ligand complex. From the Figs.\ 2(a) and (c), we obtain the fit values $k_\text{RL} = 7.2 \pm 0.7 \, k_B T/\text{nm}^2$ and $k_\text{RL} = 6 \pm 1 \, k_B T/\text{nm}^2$ for the effective spring constant of the lipid-anchored receptor-ligand complex, which are consistent within the numerical accuracy and significantly smaller than the value $k_\text{RL} = 28 \pm 2 \, k_B T/\text{nm}^2$ for the transmembrane receptor-ligand complex. The effective spring constant $k_\text{RL}$ of the lipid-anchored receptor-ligand complex thus appears to be dominated by spatial variations of the anchor points, and not by variations in the binding-site distance.  For the effective spring constant $k_\text{TS}$ of the transition-state complex, we obtain fit values between $1.5\, k_B T/\text{nm}^2$ and  $2.5\, k_B T/\text{nm}^2$ for the lipid-anchored and transmembrane receptors and ligands that agree within numerical accuracy (see Figs.\ 2(b) and (e) and 3(b) and (e)), which indicates that $k_\text{TS}$ is dominated by the identical ectodomains  of the two types of receptors and ligands.
Based on the fit results of data from the same membrane systems, the preferred length $L_0$ of the bound receptor-ligand complex appears to be identical to the preferred length $L_\text{TS}$ of the transition-state complex within the numerical accuracy. However, the fit values for the preferred lengths $L_0$ and $L_\text{TS}$ in Figs.\ 2(d) and (e)  are somewhat smaller than the fit values for these lengths in Figs.\ 2(a) and (b).

Compared to Eq.\ (18) of our accompanying article \cite{Xu}, Eq.\ (\ref{K2Dl}) includes the additional numerical factor $c_\text{2D}$ to quantify deviations between theory and MD data. The fit values $c_\text{2D}$ in Figs.\ 2(a) and (d) and Fig.\ 3(a) indicate that the general theory derived in our accompanying article \cite{Xu} overestimates $K_\text{2D}$ by a factor $1/c_\text{2D}$ between 2 and 3 compared to the MD data. In contrast, this theory is in good agreement with the $K_\text{2D}$ data from MC simulations \cite{Xu}. A difference between the MC simulations described in our accompanying article \cite{Xu} and our MD simulations is that the anchor positions of the receptors and ligands slightly fluctuate relative to the membrane midsurfaces in the MD simulations. { These slight protrusion fluctuations of the anchors in the direction perpendicular to the membranes are absent in our MC simulations}, in which the anchor points of the receptors and ligands are confined to the discretized membrane surfaces of an elastic-membrane model \cite{Xu}. However, a general theoretical investigation of protrusion fluctuations of the anchors of receptors and ligands relative to the membrane midsurfaces is beyond the scope of this article. 

\begin{figure}[t]
\includegraphics[width=0.95\columnwidth]{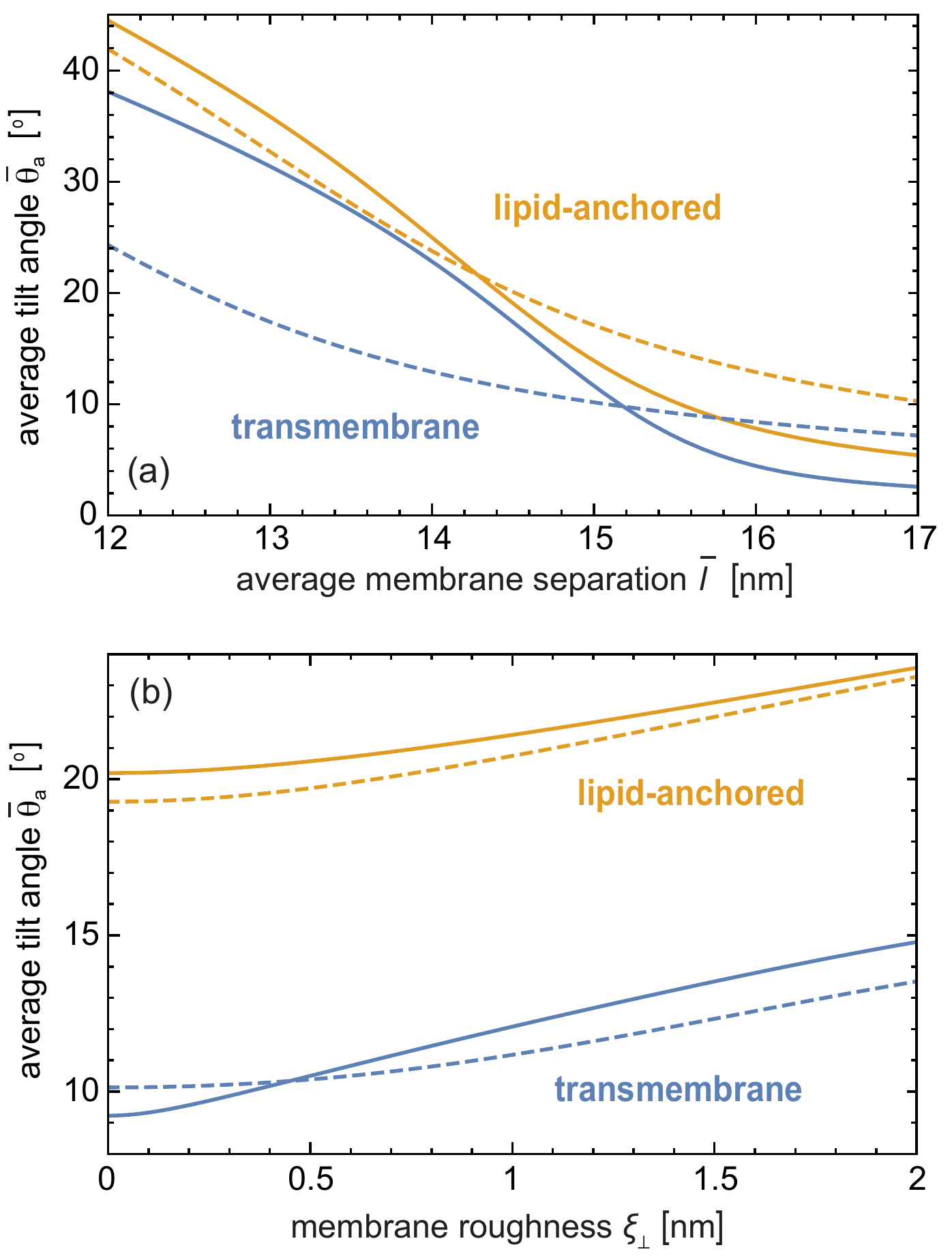}
\caption{
Average tilt angles $\bar{\theta}_a$ of the receptor-ligand complexes (full lines) and transition-state complexes (dashed lines) of the lipid-anchored and transmembrane receptors and ligands in the membrane systems of Fig.\ 2 and 3. The average tilt angles are calculated based on Eqs.\ (\ref{bartheta}) and (\ref{barthetal}) (a) for the relative membrane roughness $\xi_\perp \simeq 0.54$ nm and  $\xi_\perp \simeq 0.53$ nm of the membrane systems of Figs.\ 2(a) to (c) and 3(a) and (c) with lipid-anchored and transmembrane receptors and ligands, respectively, and (b) for the preferred average membrane separation $\bar{l} \simeq 14.1$ nm and $\bar{l} \simeq 15.1$ nm of the lipid-anchored and  transmembrane receptors and ligands, respectively, as in Figs.\ 2(d) to (f) and 3(d) to (f). In these calculations, we have used the fit values specified in Figs.\ 2 and 3. 
}
\label{figure_tilt_angles}
\end{figure}

An important element of our MD simulations and theory is the tilt of the receptors and ligands relative to the membrane normals. In our theory, the average tilt angle of the complexes can be calculated from
\begin{equation}
\bar{\theta}_a =  \int \bar{\theta}_a(l) P(l) \text{d}l
\label{bartheta}
\end{equation}
and
\begin{equation}
\bar{\theta}_a(l) = \frac{\int_0^{\pi/2} \theta_a  e^{-H(l,\theta_a)/k_B T}  \sin\theta_a \text{d}\theta_a}{\int_0^{\pi/2} e^{-H(l,\theta_a)/k_B T}  \sin\theta_a \text{d}\theta_a}
\label{barthetal}
\end{equation}
with $H(l,\theta_a) = H_\text{RL}(l,\theta_a)$ for the receptor-ligand complexes and $H(l,\theta_a) = H_\text{TS}(l,\theta_a)$ for the transition-state complexes. Fig.\ \ref{figure_tilt_angles} illustrates how the average tilt angle $\bar{\theta}_a$ of the receptor-ligand complexes (full lines) and transition-state complexes (dashed lines) depends on the average membrane separation $\bar{l}$ and relative roughness $\xi_\perp$ for the membrane systems of Fig.\ 2 and 3. In Fig.\ \ref{figure_tilt_angles}(a), the average tilt angle strongly increases for average separations $\bar{l}$ that are smaller than the preferred average separations  $\bar{l}_0\simeq 14.1$ nm and  $\bar{l}_0\simeq 15.1$ nm of the lipid-anchored and transmembrane complexes, respectively. These increased  tilt angles allow the complexes to bind at smaller separations and lead to the asymmetry of the functions $K_\text{2D}(\bar{l})$ and $k_\text{on}(\bar{l})$ relative to the preferred average separations at which $K_\text{2D}$ and $k_\text{on}$ are maximal (see Figs.\ 2(a) and (b)). At the preferred average separations for binding, the average tilt angles are about 20$^\circ$ for the lipid-anchored complexes and about 10$^\circ$ for the transmembrane complexes  for small roughnesses $\xi_\perp$, and increase with increasing roughness (see  Fig.\ \ref{figure_tilt_angles}(b)). For small roughnesses $\xi_\perp$, the tilt of the complexes at the preferred average membrane separations for binding results from the fact that the density of conformations accessible to the complexes is proportional to $\sin \theta_a$. The rotational entropy of the complexes therefore increases with the  tilt angle $\theta_a$, and the average tilt angles of the lipid-anchored and transmembrane complexes follow from the interplay of this rotational entropy and the energetic penalty $k_a \theta_a^2$ for tilting. For larger roughnesses, the average tilt angles $\bar{\theta}_a$ increase because the broader distributions $P(l)$ than include local membrane separations $l$ that are significantly smaller than the preferred separations and induce a stronger tilt of the complexes. The average tilt angles of the unbound receptors and ligands are $\bar{\theta}_a = \int \theta_a  e^{-k_a \theta_a^2/2k_B T}  \sin\theta_a \text{d}\theta_a/ \int e^{-k_a \theta_a^2/2k_B T}  \sin\theta_a \text{d}\theta_a \simeq 41^\circ$ for the lipid-anchored receptors and ligands with anchoring strength $k_a \simeq 2.5 \, k_B T$, and $\bar{\theta}_a \simeq 15^\circ$ for the transmembrane receptors and ligands with anchoring strength $k_a \simeq 23\, k_B T$.

\section{Discussion and Conclusions}

In this article, we have presented a general theory for the binding kinetics of membrane-anchored receptors and ligands, and have corroborated this theory by a comparison to detailed data from MD simulations with lipid-anchored and transmembrane receptors and ligands. The theory describes how the binding rate constants $k_\text{on}$ and  $k_\text{off}$ and binding equilibrium constant $K_\text{2D}$ depend (i) on characteristic parameters of the membrane-anchored receptor and ligand molecules, and (ii) on the average membrane separation $\bar{l}$ and relative roughness $\xi_\perp$ of the membranes. For relative membrane roughnesses $\xi_\perp$ that are much larger than the widths $\xi_\text{RL}$ and $\xi_\text{TS}$ of the functions $K_\text{2D}(l)$ and $k_\text{on}(l)$ introduced in Eqs.\ (\ref{K2Dl}) and (\ref{konl}), the binding equilibrium constant $K_\text{2D}$ and binding on-rate constant $k_\text{on}$ are inversely proportional to  $\xi_\perp$ at the preferred average membrane separations $\bar{l}_0$ and  $\bar{l}_\text{TS}$ of the receptor-ligand and transition-state complexes (see Eqs.\ (\ref{K2Dlim}) and (\ref{konlim})). For such large roughnesses, the off-rate constant $k_\text{off}$ is independent both of the relative roughness and average separation of the membranes if the preferred average membrane separation $\bar{l}_\text{TS}$ of the transition-state complexes is approximately equal to the preferred average separation $\bar{l}_0$ of the bound receptor-ligand complexes, which appears to be the case for our lipid-anchored and transmembrane receptors and ligands (see fit parameters in Figs.\ 2 and 3 and previous section). Based on expressions given below Eqs.\ (\ref{K2Dlim}) and (\ref{konlim}), the widths of the functions $K_\text{2D}(l)$ and $k_\text{on}(l)$ can be estimated as $\xi_\text{RL} \simeq 2.1$ nm and $\xi_\text{TS} \simeq 2.2$ nm for our lipid-anchored receptors and ligands, and $\xi_\text{RL} \simeq 0.38$ nm and $\xi_\text{TS} \simeq 0.8$ nm for our transmembrane receptors and ligands

{
The dependence of the binding constant $K_\text{2D}$ and rate constants $k_\text{on}$ and $k_\text{off}$ on the average membrane separation $\bar{l}$ and relative membrane roughness $\xi_\perp$ leads to a cooperative binding of membrane-anchored receptors and ligands. The formation of receptor-ligands bonds facilitates the binding of additional receptors and ligands because these bonds (i) 'adjust' the average separation of adhering membrane segments to values that are closer or identical to the preferred average separation for binding, and (ii) reduce the relative roughness of the membrane segments by constraining membrane shape fluctuations. 
Membrane adhesion zones are typically large compared to the average distance of about $1/\sqrt{[\text{RL}]}$ between neighboring receptor-ligand bonds. If the adhesion is dominated by a single species of receptors R and ligands L, the average membrane separation is close to the preferred average separation for binding of these receptors and ligands. The relative roughness of the adhering membranes is then proportional to the average distance between neighboring receptor-ligand bonds \cite{Krobath07,Xu}:
\begin{equation}
\xi_\perp \propto 1/\sqrt{[\text{RL}]}
\label{xiperp}
\end{equation}
For roughness values $\xi_\perp$ larger than the widths $\xi_\text{RL}$ and $\xi_\text{TS}$ of the functions $K_\text{2D}(l)$ and $k_\text{on}(l)$,  the binding constant $K_\text{2D}$ and on-rate constant $k_\text{on}$ in turn are inversely proportional to $\xi_\perp$ (see paragraph above). With Eq.\ (\ref{xiperp}), this inverse proportionality leads to the quadratic relation \cite{Krobath09,Hu13}
\begin{equation}
[\text{RL}] = K_\text{2D} [\text{R}][\text{L}]\propto [\text{R}]^2[\text{L}]^2
\label{loma}
\end{equation}
between the bond concentration $[\text{RL}]$ and the concentrations $[\text{R}]$ and $[\text{L}]$ of the unbound receptors and ligands, and to an overall reaction rate
\begin{equation}
r_\text{on} = k_\text{on} [\text{R}][\text{L}] \propto [\text{R}]^2[\text{L}]^2
\end{equation}
that depends quadratically on the concentrations $[\text{R}]$ and $[\text{L}]$ of the unbound receptors and ligands. These quadratic laws of mass action reflect the cooperative binding of membrane-anchored receptors and ligands.}

Because the binding interactions are identical for our lipid-anchored and transmembrane receptors and ligands, the substantially smaller $K_\text{2D}$ values of the lipid-anchored receptors and ligands at average membrane separations around $\bar{l}_0$ result from the softer anchoring compared to the transmembrane receptors and ligands. In our theory, the effect of the anchoring on $K_\text{2D}$ is reflected in the rotational phase space volumes $\Omega_\text{R}$, $\Omega_\text{L}$, and $\Omega_\text{RL}(l)$ of the unbound receptors, unbound ligands, and bound receptor-ligand complexes, which determine the rotational free energies of the unbound receptors and ligands and bound complexes. The smaller $K_\text{2D}$ values of the lipid-anchored receptors and ligands result from a larger loss in rotational free energy upon binding due to the softer anchoring, compared to the transmembrane receptors and ligands.

An important element in our theory is the tilt of the receptor-ligand and transition-state complexes relative to the membrane normals. The tilt of the complexes is reflected by effective configurational energies $H_\text{RL}(l,\theta_a)$ and  $H_\text{TS}(l,\theta_a)$ of the receptor-ligand and transition-state complexes in Eqs.\ (\ref{HRL}) and (\ref{HTS}) that depend both on the local separation $l$ of the membranes and on the tilt angle $\theta_a$ of the complexes. The calculation of the functions $K_\text{2D}(l)$ and $k_\text{on}(l)$ in Eq.\ (\ref{K2Dl}) with Eq.\ (\ref{OmegaRL}) and Eq.\ (\ref{konl}) is based on these effective configurational energies and involves integrations over the  tilt angle $\theta_a$. Our functions  $K_\text{2D}(l)$ and $k_\text{on}(l)$ are asymmetric with respect to the locations $l_0$ and $l_\text{TS}$ of their maxima because the tilting facilitates the formation of complexes at separations smaller than $l_0$ and $l_\text{TS}$, compared to larger separations. This asymmetry of the functions  $K_\text{2D}(l)$ and $k_\text{on}(l)$ explains the asymmetry in the MD data of Figs.\ 2(a) and (b). The asymmetry of  $K_\text{2D}(l)$ for rod-like and semi-flexible receptors and ligands is also directly corroborated by Monte Carlo data in our accompanying article \cite{Xu}. In previous theories\cite{Dembo88,Gao11,Bihr12}, the effective configurational energies of membrane-anchored receptor-ligand and transition-state complexes depend only on the membrane separation $l$. In harmonic approximation, such effective configurational energies $H_\text{RL}(l) = \frac{1}{2}K_\text{RL} (l - l_0)^2$ and $H_\text{TS}(l) = \frac{1}{2}K_\text{TS} (l - l_\text{TS})^2$ lead to symmetric Gaussian functions  $K_\text{2D}(l) \sim \exp(-\frac{1}{2}K_\text{RL} (l - l_0)^2/k_B T)$ and $k_\text{on}(l) \sim \exp(-\frac{1}{2}K_\text{TS} (l - l_\text{TS})^2/k_B T)$, in contrast to our theory and simulation data.

\section*{Acknowledgements}

The authors thank the Deutsche Forschungsgemeinschaft (DFG) for financial support {\em via} the International Research Training Group 1524 ``Self-Assembled Soft Matter Nano-Structures at Interfaces".

\appendix

%
\section{Coarse-grained molecular model of biomembrane adhesion}

Our coarse-grained model includes water, lipid molecules, and receptor and ligand molecules. Water molecules (W) are represented by single beads. 

{\em Lipids}.---A lipid molecule consists of three hydrophilic head beads (H) and two hydrophobic chains (C) with four beads each \cite{Goetz98,Shillcock02,Grafmuller07, Grafmuller09}. Adjacent beads are connected via harmonic potentials
\begin{equation}
V_\text{bond}(r) =  \frac{1}{2}\,k_r\left(r-l_0\right)^2
\label{Eq:HarmonicPotential}
\end{equation}
with bond strength $k_r=128\,k_BT/r_0^2$ and preferred bond length $l_0 = 0.5\,r_0$ \cite{Shillcock02}. Here, $r$ is the distance between the two beads, and $r_0$ is the characteristic length of the beads. The two hydrophobic chains of the lipid molecules are stiffened by the bending potential \cite{Goetz98}
\begin{equation}
V_\text{bend}(\phi) = k_\phi\left[1-\cos (\phi-\phi_0)\right]
\label{Eq:BendingPotential}
\end{equation}
that acts between two consecutive bonds along each chain. The bending constant is $k_\phi = 15\, k_BT$, and the bond angle $\phi$ attains the preferred value $\phi_0=0$ for collinear bonds.

{\em Transmembrane receptors and ligands}.---Our transmembrane receptor and ligand molecules consist of a transmembrane domain and an ectodomain. The transmembrane domain is composed of four layers of lipid-chain-like beads (T$_\text{C}$), which are shown in yellow in Fig. 1, in between two layers of lipid-head-like beads (T$_\text{H}$) shown in blue. The ectodomain consists of six layers of a hydrophilic bead type I. Each pair of nearest neighboring beads of a receptor or ligand is connected by a harmonic potential with bond strength $k_r = 128\,k_BT/r_0^2$ and preferred bond length $l_0 = 0.875\,r_0$. This bond length corresponds to the average distance of neighboring water beads in our simulations with bead density $\rho = 3\,r_0^{-3}$. Each pair of next-nearest neighboring beads in two adjacent layers of a receptor or ligand is connected by a harmonic potential with $k_r = 128\,k_BT/r_0^2$ and $l_0 = \sqrt{2}\times 0.875\,r_0$.

{\em Lipid-anchored receptors and ligands}.---The ectodomain of our lipid-anchored receptor and ligand molecules is identical to the ectodomain of our transmembrane receptors and ligands. This ectodomain is directly anchored to a lipid molecule. 
 The central bead in the bottom layer of the ectodomain is connected by a harmonic potential (\ref{Eq:HarmonicPotential}) with $k_r = 128\,k_BT/r_0^2$ and $l_0 = 0.875\,r_0$ to the first head bead of the lipid. This head bead is not connected to either chain of the lipid.

{\em Nonbonded interactions}.---In addition to the forces arising from the bonded interactions specified above, each pair of beads--except for the interaction beads of a receptor and a ligand (see below)--exhibit the soft repulsive forces
\begin{equation}
{\bf F}_{ij} = \left\{
\begin{array}{lr}
a_{ij}(1-r_{ij}/r_0)\hat{{\bf r}}_{ij}, & r_{ij} < r_0\\
0, & r_{ij} \ge r_0\\
\end{array}
\right.
\label{Eq:ConservForce}
\end{equation}
with a repulsion strength $a_{ij}$ that depends on the types of the two beads $i$ and $j$. The different repulsion strengths shown in Table \ref{Tab:S1} reflect the chemical nature of the beads, i.e.\ their hydrophilicity or hydrophobicity. To avoid a clustering of receptors and ligands, the repulsion strength between the beads of two different receptors or two different ligands adopts the value $a_{ij}=75\,k_BT/r_0$, which is larger than the repulsion strength $a_{ij}=25\,k_BT/r_0$ between two beads of the same receptor, of the same ligand, or of a receptor and a ligand. 

\begin{table}[b]
\caption{{\label{Tab:S1}}Repulsion strengths $a_{ij}$ in units of $k_BT/r_0$. Numbers in parentheses indicate the repulsion strength between the beads of two different receptors or two different ligands.}
\begin{tabular}{l|cccccc}
$a_{ij}$ & W & H & C & T$_{\rm H}$ & T$_{\rm C}$ & I \\
\hline
W & 25 & 30 & 75 & 30 & 75 & 25 \\ 
H & 30 & 30 & 35 & 30 & 35 & 30 \\ 
C & 75 & 35 & 10 & 35 & 10 & 35 \\ 
T$_{\rm H}$ & 30 & 30 & 35 & 25(75) & 25(75) & 25(75) \\ 
T$_{\rm C}$ & 75 & 35 & 10 & 25(75) & 25(75) & 25(75) \\ 
I & 25 & 30 & 35 & 25(75) & 25(75) & 25(75) \\ 
\end{tabular}
\end{table}

{\em Specific binding of receptors and ligands}.---
The binding of a receptor and a ligand is modeled via the binding potential \cite{Hu13}
\begin{equation}
V_\text{bind}(r,\,\theta) = v_\text{bind}(r)\, e^{-k_\theta(\theta-\theta_0)^2}
\label{Eq:BindingPotential}
\end{equation}
which depends both on the distance $r$ between the interaction beads of the receptor and ligand and on the angle $\theta$ between the two molecules. The interaction beads are located in the center of the top layer of the ectodomains and are indicated in black in Fig. 1. The angle $\theta$ between a receptor and a ligand is defined as the angle between the two bonds that connect the interaction beads of the molecules to the central beads of the adjacent bead layers. The distance-dependent term $v_{\rm bind}(r)$ of the specific interaction is
\begin{align}
v_\text{bind}(r) = \left\{
\begin{array}{lr}
\frac{1}{2}r_0\left[a_{\rm II}(1-r/r_0)^2-F_m\right], & r < r_0\\
F_mr_0\left[(1-r/r_0)^2-\frac{1}{2}\right], &\hspace*{-0.5cm} r_0 \le r < \frac{3}{2}r_0\\
-F_mr_0(2-r/r_0)^2, &\hspace*{-1cm} \frac{3}{2}r_0 \le r < 2r_0\\
0, & r \ge 2r_0\\
\end{array}
\right.
\label{Eq:BindingPotentialRadialPart}
\end{align}
with repulsion strength $a_{\rm II} = 25\,k_BT/r_0$ and the attraction strength $F_m = 16\,k_BT/r_0$. Differentiating $v_{\rm bind}(r)$ with respect to $r$ leads to the radial force component
\begin{equation}
F_\text{bind}(r) = \left\{
\begin{array}{lr}
a_{\rm II}(1-r/r_0), & r < r_0\\
2F_m(1-r/r_0), & r_0 \le r < \frac{3}{2}r_0\\
2F_m(r/r_0-2), & \frac{3}{2}r_0 \le r < 2r_0\\
0, & r \ge 2r_0\\
\end{array}
\right.
\label{Eq:BindingForceRadialPart}
\end{equation}
which includes a soft repulsion for $r < r_0$ and an attraction for $r_0 \le r < 2 r_0$. The parameter $k_\theta$ in Eq. (\ref{Eq:BindingPotential}) determines the width of the binding-angle distribution and is chosen to be $k_\theta = 10/{\rm rad}^2$. The angle $\theta$ adopts the preferred value $\theta_0=0$ if the two molecules are facing each other. The binding potential attains its minimum value of $-\frac{1}{2}F_mr_0 = -8\,k_BT$ at $r=r_0$ and $\theta=\theta_0$. For this intermediate binding energy of the receptors and ligands, both stable bonds and a large number of binding and unbinding events can be observed in our simulations. 

\section{MD simulations}

We have performed simulations with dissipative particle dynamics (DPD), a coarse-grained MD technique that explicitly includes water and reproduces the correct hydrodynamics  \cite{Hoogerbrugge92, Espanol95, Groot97}.

{\em Time step and initial relaxation}.---In our simulations, Newton's equations of motion are numerically integrated with a time step $t_0 = 0.03\sqrt{m_0r_0^2/k_BT}$ using the velocity-Verlet algorithm as in ref.\ \cite{Groot97}. Here, $m_0$ is the mass of the beads, and $k_B T$ is the thermal energy. For this time step, the average temperature of the beads deviates from the expected value by at most 2\%. Our optimized DPD code is parallelized to achieve a speedup of about a factor 6 by using 8 CPU cores, which enable us to simulate up to tens of thousands of binding and unbinding events to determine the binding rate constants with high accuracy. A relaxation run of $2-5\cdot 10^6\,t_0$ is performed for thermal equilibration in each system before statistical sampling.

{\em Length and time scales}.---Physical length and time scales for the characteristic bead length $r_0$ and for the time step $t_0$ of our simulations can be obtained from a comparison to experimental data for dimyristoyl-phosphatidylcholine (DMPC) fluid bilayers. Our model lipids correspond to the phospholipid DMPC since each C bead of the lipid tails can be seen to represent 3.5 CH$_2$ groups \cite{Goetz98,Grafmuller07}, which leads to a total tail length of 14 CH$_2$ groups as for DMPC. From the experimental value $d_{\rm HH}\simeq 3.53$ nm for the vertical distance between the head groups of two DMPC monolayers in fluid bilayers \cite{Kucerka05} and our simulation result $d_{\rm HH}\simeq 3.64\,r_0$, we obtain the physical length scale $r_0\simeq 1.0$ nm. From the experimentally measured lateral diffusion coefficient $D\simeq 5$ $\mu$m$^2$/s of DMPC \cite{Oradd02}  and our simulation result $D\simeq 5.7\cdot 10^{-4}\,r_0^2/t_0$, we obtain the physical time scale $t_0 \simeq 114$ ns.  A characteristic length of our binding potential is the standard deviation of the interaction beads in the direction parallel to a bound receptor-ligand complex.  From our simulations, we obtain the value 0.19 nm for this characteristic width of the binding potential \cite{Hu13}. The standard deviation of the binding angle $\theta$ of receptor-ligand complexes in our simulations is 0.084.

{\em Simulation box}.---The lipid membranes in our simulations are confined within a simulation box with size $V= L_x\times L_y\times L_z$ and periodic boundary conditions. The box extension in the direction perpendicular to the membranes has the same value $L_z = 40$ nm in all our simulations, while the extensions $L_x=L_y$ are varied to simulate different membrane sizes.  The number density of beads in the simulation box is set to $\rho = 3\,r_0^{-3}$. The total number of beads thus is $\rho V$. In our simulations without confining membrane potentials, the number of lipids is adjusted such that the membrane tension vanishes. Our smallest membranes have an area of $14\times 14$ nm$^2$ and contain 296 lipids and single receptor and ligand molecules. Our largest membranes have an area of $120\times 120$ nm$^2$ and are composed of 22,136 lipids and 25 transmembrane receptors and ligands. The average separation between the membranes is kept constant in our simulations, while the local separations fluctuate around the average value with a normal distribution as shown in Fig. \ref{figure_Pl}. 

{\em Confining membrane potentials}.---In our simulations with confining potentials, we impose an additional harmonic potential
\begin{equation}
V_{\rm conf}(z) = \frac{1}{2}k_{\rm conf}(z-z_0)^2
\end{equation}
on one of the three head beads of each lipid in the two distal monolayers of the apposing membranes, i.e.\ in the two monolayers that do not face the other membrane. The potential is acting on the head bead connected to the right side chain of the lipid molecule shown in Fig.\ 1. The $z$-direction of our simulation box is on average perpendicular to the membranes. In Figs. 2(d)-2(f), the red data points are from simulations with the confining strengths $k_{\rm conf} = 64$, $4$, $2$ and $1$ $k_BT$/nm$^2$ (from left to right). The membrane tensions for these confining strengths are $-0.12\pm 0.03$, $-0.02\pm 0.01$, $0.06\pm 0.01$ and $0.19\pm 0.01$ $k_BT$/nm$^2$, respectively. 

\section{Analysis of binding kinetics}

Binding and unbinding events of receptor and ligand molecules in our simulations can be identified from the distance $r$ between the interaction beads of the molecules  \cite{Hu13}. A binding event is defined to occur when $r$ falls below the binding threshold $r_1 = 1$ nm, at which the binding potential in Eq. (\ref{Eq:BindingPotential}) attains its minimum. An unbinding event is defined to occur when $r$ exceeds the unbinding threshold $r_2 = 4$ nm, which is well beyond the range of fluctuations in the bound state \cite{Hu13}. Our values for the binding equilibrium constant $K_\text{2D}$ and the relative values of on-rate constants $k_\text{on}$ and off-rate constants $k_\text{off}$ do not depend on the choice of these thresholds.

The binding and unbinding events divide our simulation trajectories into different states with different numbers of receptor-ligand complexes. In our simulations with $N_R$ receptors and $N_L$ ligands, we have $N+1$ states where $N=\min{(N_R,\,N_L)}$ is the maximum number of complexes. Our simulation trajectories can be mapped to a Markov model
\begin{equation}
0\xrightleftharpoons[k_{-}^{(1)}]{k_{+}^{(0)}}1\xrightleftharpoons[k_{-}^{(2)}]{k_{+}^{(1)}}2\xrightleftharpoons[k_{-}^{(3)}]{k_{+}^{(2)}}3\cdots N-1\xrightleftharpoons[k_{-}^{(N)}]{k_{+}^{(N-1)}}N
\end{equation}
with binding rate $k_{+}^{(n)} = (1/A)(N_L - n)(N_R - n)k_\text{on}^{(n)}$ for the transition from state $n$ to $n+1$ and unbinding rate $k_{-}^{(n)} = n k_\text{off}^{(n)}$ for the transition from state $n$ to $n-1$. Here, $A$ is the membrane area, $k_\text{on}^{(n)}$ and $k_\text{off}^{(n)}$ are the on- and off-rate constants in state $n$. By maximizing the likelihood function for our whole simulation trajectory, we obtain the maximum likelihood estimators for the rate constants
\begin{equation}
k_\text{on}^{(n)}=\cfrac{N_n^+A}{(N_R-n)(N_L-n)T_n} \quad \text{and}\quad k_\text{off}^{(n)}=\cfrac{N_n^-}{nT_n}
\end{equation}
where $N_n^+$ is the total number of transitions from $n$ to $n+1$, $N_n^-$ the total number of transitions from $n$ to $n-1$, and $T_n$ the total dwell time in state $n$. Our estimator for the binding constant then is \cite{Hu13}
\begin{equation}
K_\text{2D}^{(n)}=\cfrac{k_\text{on}^{(n-1)}}{k_\text{off}^{(n)}}=\cfrac{nAT_n}{(N_R-n+1)(N_L-n+1)T_{n-1}}
\end{equation}
because the transition numbers $N_{n-1}^+$ and $N_n^-$ are identical in equilibrium.\\


%
%

%



%

\end{document}